\documentclass[twoside,lettersize,ieeeconf]{IEEEtran}
\usepackage{amsmath,amsfonts}
\usepackage{algpseudocode} 
\usepackage{array}
\usepackage[caption=false,font=normalsize,labelfont=sf,textfont=sf]{subfig}
\usepackage{textcomp}
\usepackage{stfloats}
\usepackage{url}
\usepackage{color}
\usepackage{verbatim}
\usepackage{graphicx}
\usepackage{multirow}
\usepackage{balance}
\usepackage{tabularx}
\usepackage{hyperref}
\usepackage{academicons}
\usepackage{xcolor}
\usepackage{svg}
\usepackage[noadjust]{cite}
\usepackage[T1]{fontenc}
\usepackage{enumitem}
\usepackage{tablefootnote}
\usepackage{caption}
\usepackage{tikz}
\usepackage{amssymb}
\usepackage{threeparttable}

\usepackage[ruled,vlined]{algorithm2e}
\def\BibTeX{{\rm B\kern-.05em{\sc i\kern-.025em b}\kern-.08em
    T\kern-.1667em\lower.7ex\hbox{E}\kern-.125emX}}

\newcommand{\modelaccuracytable}{
\begin{table}[ht]
\centering
\captionsetup{justification=centering}
\caption{Accuracy of classifier Models.}
\label{table:1}
\begin{tabular}{|c|c|}
\hline
Model                      & Test Accuracy \\ \hline
Single Input (MFCC)        &          97.45     \\ \hline
Dual Input (MFCC + LogMel) &      99.63         \\ \hline
\end{tabular}
\end{table}
}

\newcommand{\finalresults}{
\begin{table}[]
\centering
\captionsetup{justification=centering}
\caption{Average accuracies of dual input classifier model with continual learning.}
\label{table:2}
\begin{tabular}{|c|ccccc|}
\hline
\multirow{2}{*}{Adaptation} & \multicolumn{5}{c|}{SNR levels}                                                                                            \\ \cline{2-6} 
                            & \multicolumn{1}{c|}{-10 dB} & \multicolumn{1}{c|}{-5 dB} & \multicolumn{1}{c|}{0 dB}  & \multicolumn{1}{c|}{5 dB}  & 10 dB \\ \hline
DWASHING                    & \multicolumn{1}{c|}{93.84}  & \multicolumn{1}{c|}{94.88} & \multicolumn{1}{c|}{95.62} & \multicolumn{1}{c|}{95.22} & 95.2  \\ \hline
NFIELD                      & \multicolumn{1}{c|}{91.44}  & \multicolumn{1}{c|}{92.56} & \multicolumn{1}{c|}{94.91} & \multicolumn{1}{c|}{96.22} & 95.09 \\ \hline
OOFFICE                     & \multicolumn{1}{c|}{92.94}  & \multicolumn{1}{c|}{92.50} & \multicolumn{1}{c|}{94.34} & \multicolumn{1}{c|}{94.94} & 94.31 \\ \hline
TCAR                        & \multicolumn{1}{c|}{94.56}  & \multicolumn{1}{c|}{95.25} & \multicolumn{1}{c|}{95.28} & \multicolumn{1}{c|}{94.78} & 95.22 \\ \hline
\end{tabular}
\end{table}
}

\newcommand{\componentsaccuracies}{
\begin{table}[]
\begin{threeparttable}
\centering
\captionsetup{justification=centering}
\caption{Accuracies obtained on OOFFICE runtime using dual input classifier model.}
\label{table:3}
\begin{tabular}{|c|c|c|ccccc|}
\hline
\multirow{2}{*}{Retrain} & \multirow{2}{*}{\begin{tabular}[c]{@{}c@{}}WDN$^1$ \end{tabular}} & \multirow{2}{*}{\begin{tabular}[c]{@{}c@{}}SDN$^2$ \end{tabular}} & \multicolumn{5}{c|}{SNR levels (dB)}                                                                                            \\ \cline{4-8} 
                            &                                                                              &                                                                                 & \multicolumn{1}{c|}{-10} & \multicolumn{1}{c|}{-5} & \multicolumn{1}{c|}{0}  & \multicolumn{1}{c|}{5}  & 10 \\ \hline
No                          & No                                                                           & Yes                                                                             & \multicolumn{1}{c|}{88.62}  & \multicolumn{1}{c|}{94.5}  & \multicolumn{1}{c|}{96.72} & \multicolumn{1}{c|}{97.97} & 98.03 \\ \hline
No                          & Yes                                                                          & Yes                                                                             & \multicolumn{1}{c|}{86.38}  & \multicolumn{1}{c|}{92.16} & \multicolumn{1}{c|}{95.47} & \multicolumn{1}{c|}{95.69} & 96.41 \\ \hline
Yes                         & No                                                                           & Yes                                                                             & \multicolumn{1}{c|}{92.88}  & \multicolumn{1}{c|}{94.53} & \multicolumn{1}{c|}{96.59} & \multicolumn{1}{c|}{96.31} & 97.03 \\ \hline
Yes                         & Yes                                                                          & Yes                                                                             & \multicolumn{1}{c|}{92.94}  & \multicolumn{1}{c|}{92.50} & \multicolumn{1}{c|}{94.34} & \multicolumn{1}{c|}{94.94} & 94.31 \\ \hline 
\end{tabular}

\begin{tablenotes}
\item[1] WDN: Wavelet Denoising 
\item[2] SDN: Spectral Denoising
\end{tablenotes}
\end{threeparttable}
\end{table}
}
\newcommand{\figureoverviewCL}{
\begin{figure*}[ht]
\captionsetup{justification=centering, belowskip=-10pt}
\centering
\resizebox{\linewidth}{!}{
\label{figure_overviewCL}
\includegraphics[]{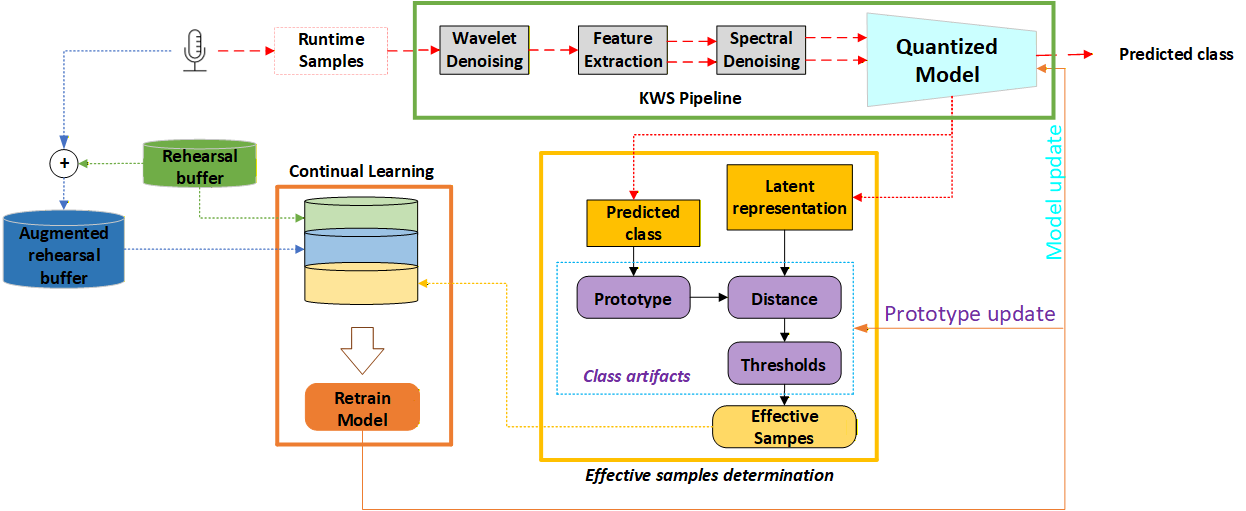}
}
\caption{Overview of proposed framework for continual learning.}
\end{figure*}
}

\newcommand{\figureeffectivesamples}{
\begin{figure}[ht]
\captionsetup{justification=centering, belowskip=-10pt}
\centering
\resizebox{\linewidth}{!}{
\label{figure_effectivesamples}
\includegraphics[]{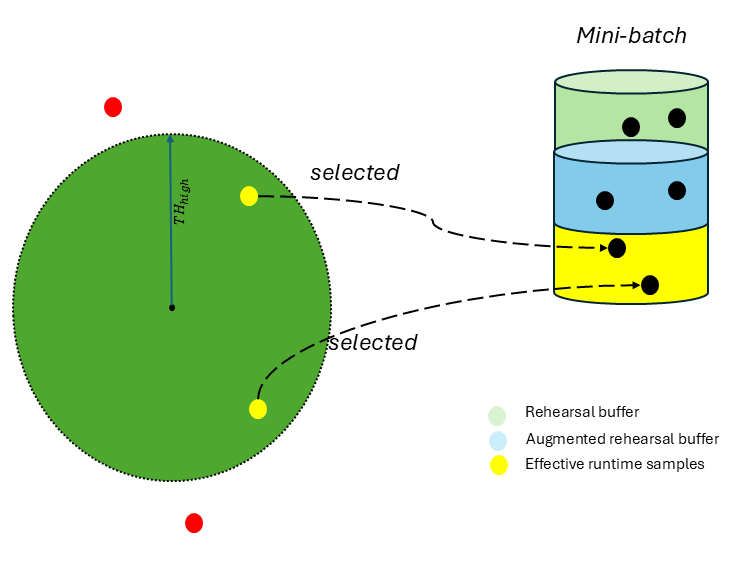}
}
\caption{Determination of effective samples to be included in training mini-batch for continual learning.}
\end{figure}
}

\newcommand{\figuresinglefeaturemodel}{
\begin{figure}[ht]
\captionsetup{justification=centering, belowskip=-10pt}
\centering
\resizebox{\linewidth}{!}{
\label{figure_singlemodelarchitecture}
\includegraphics[]{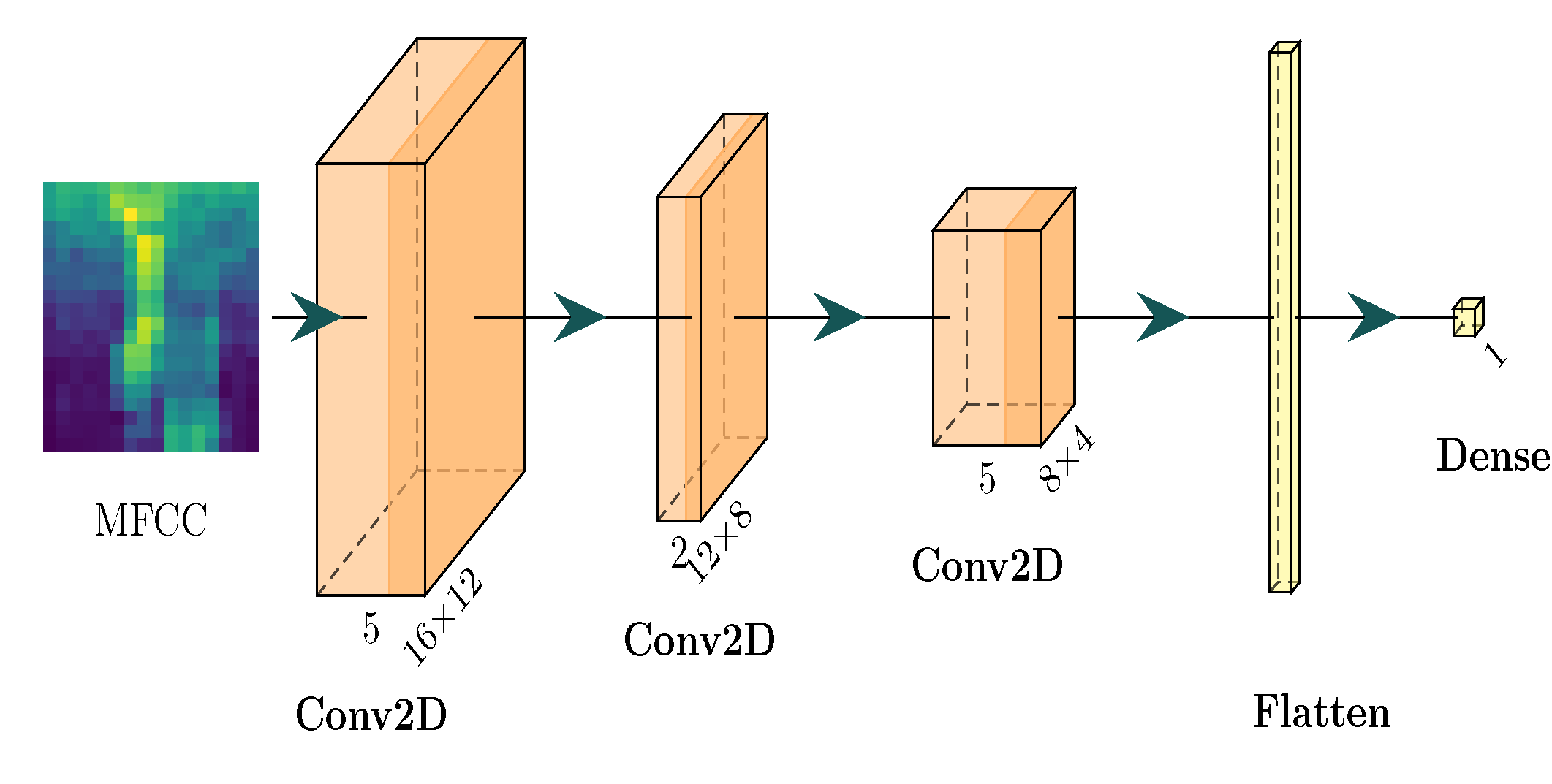}
}
\caption{Proposed classifier model (single input) architecture for continual learning.}
\end{figure}
}

\newcommand{\figuredualfeaturemodel}{
\begin{figure}[ht]
\captionsetup{justification=centering, belowskip=-10pt}
\centering
\resizebox{\linewidth}{!}{
\label{figure_dualmodelarchitecture}
\includegraphics[]{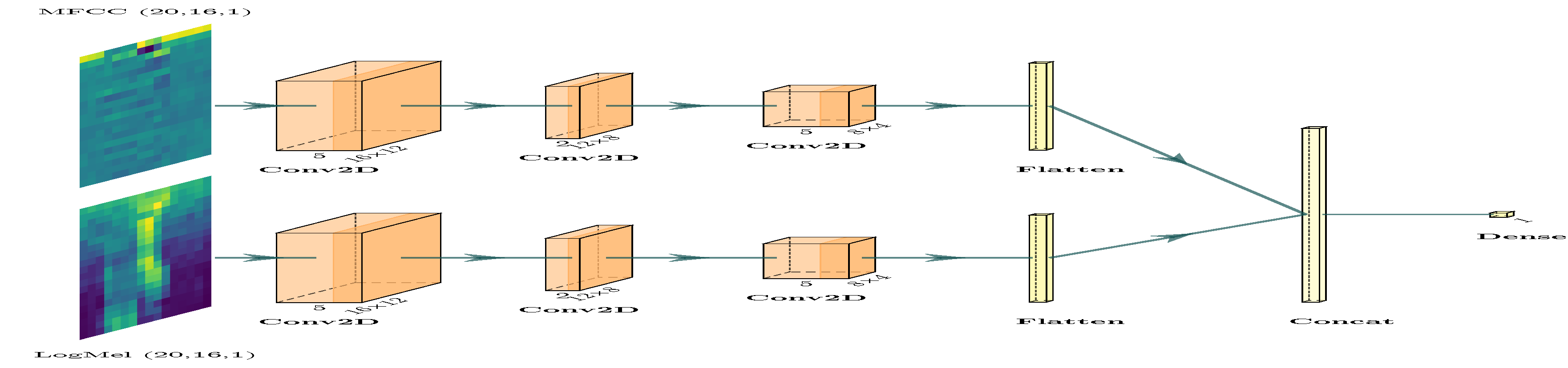}
}
\caption{Proposed classifier model (dual input) architecture for continual learning.}
\end{figure}
}

\newcommand{\figuredenoising}{
\begin{figure}[ht]
\captionsetup{justification=centering, belowskip=-10pt}
\centering
\resizebox{\linewidth}{!}{
\label{figure_denoising}
\includegraphics[]{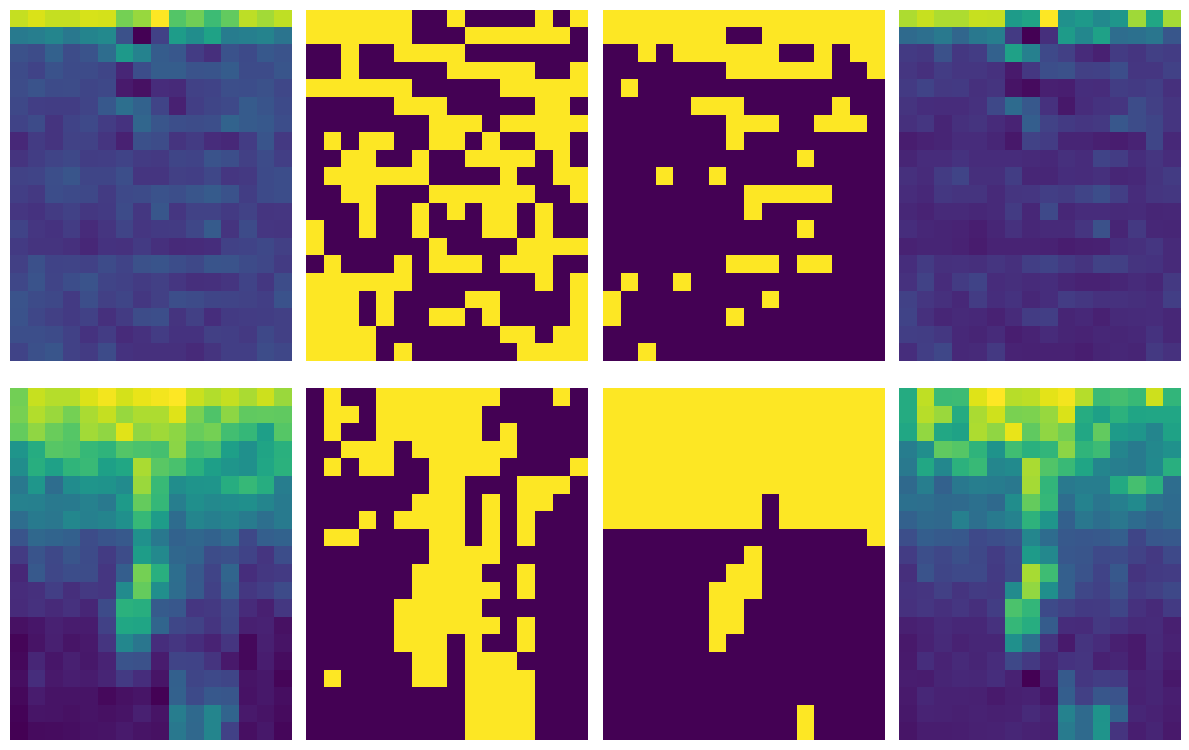}
}
\caption{An example of denoising masks generated for MFCC (top row) and Mel-Spectrogaram (bottom row).}
\end{figure}
}

\newcommand{\figureablationalpha}{
\begin{figure*}[ht]
\captionsetup{justification=centering, belowskip=-10pt}
\centering
\resizebox{\linewidth}{!}{
\label{figure_ablationalpha}
\includegraphics[]{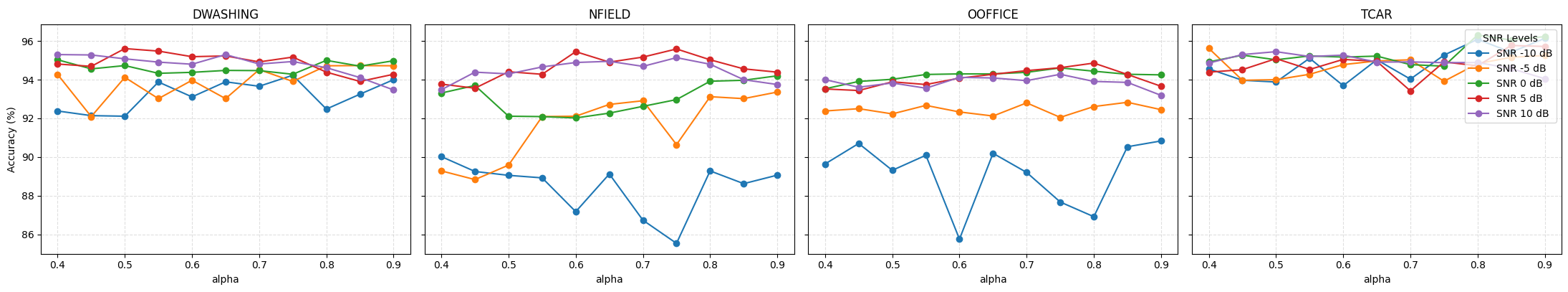}
}
\caption{Sensitivity analysis of attenuation coefficient $\alpha$ (refer Equation 7) in spectral denoising step.}
\end{figure*}
}

\newcommand{\figureablationthprob}{
\begin{figure*}[ht]
\captionsetup{justification=centering, belowskip=-10pt}
\centering
\resizebox{\linewidth}{!}{
\label{figure_ablationthprob}
\includegraphics[]{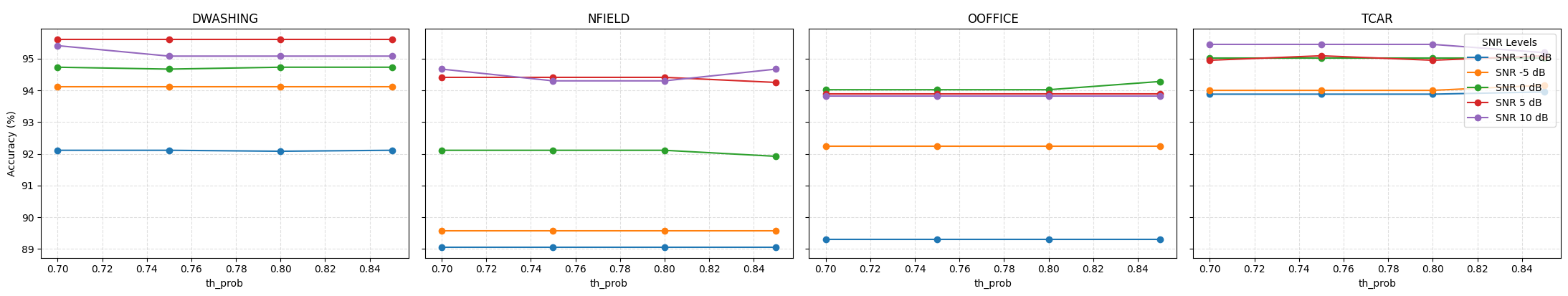}
}
\caption{Analysis of impact of classification confidence threshold $P_{th}$ (refer Algorithm I - Line 4 ) during effective sample determination.}
\end{figure*}
}

\newcommand{\figureablationthdist}{
\begin{figure*}[ht]
\captionsetup{justification=centering, belowskip=-10pt}
\centering
\resizebox{\linewidth}{!}{
\label{figurea_blationthdist}
\includegraphics[]{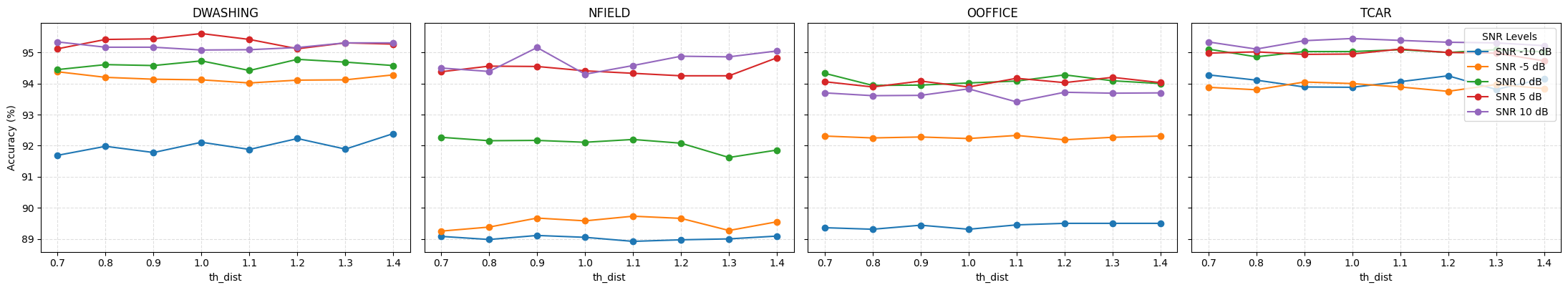}
}
\caption{Analysis of impact of distance threshold from prototype $P_{th}$ (refer Algorithm I - Line 7 ) during effective sample determination.}
\end{figure*}
}

\newcommand{\algorithmone}{%Algorithm I
\begin{algorithm}[h]
\label{algorithm_one}
\KwIn{$\{P_{th}\}$ Threshold for predicted class probability, $\{d_{th}\}$ Threshold for class prototype distance, and  $\{P_{class}\}$ Class prototypes}
\KwOut{ Mini-batch of training samples for continual learning ($M_{CL}$)}

\nl for $t = 0, 1, ..., t_n$:

{\fontfamily{qcr} \selectfont //runtime samples} \\
\nl \hskip0.25em $X_i$ $\forall i \epsilon [0, n]$ 

{\fontfamily{qcr} \selectfont //Quantized inference} \\
\nl \hskip0.25em $Y_i \leftarrow f_h(f_\theta(X_i))$   

{\fontfamily{qcr} \selectfont //Thresholding based on confidence } \\
\nl \hskip0.25em if $(max(Y_i) > P_{th})$   

\nl \hskip0.5em $class \leftarrow index(max(y_i))$ 

\nl \hskip0.5em $ d_p \leftarrow (P_{class} - f_\theta(X_i)) $

{\fontfamily{qcr} \selectfont //Thresholding based on similarity} \\
\nl \hskip0.5em if $MAE : (d_{th} > |d_p - \mu|)$  

{\fontfamily{qcr} \selectfont //save effective samples}\\
\nl \hskip0.75em if $M \leftarrow \{X_i,Y_i\} $ 

\caption{{\bf Algorithm for determining effective samples} \label{Inference}}

\end{algorithm}
}

\newcommand{\algorithmtwo}{%Algorithm II
\begin{algorithm}[h]
\label{algorithm_two}
\KwIn{$\{M_{CL} \epsilon \{X_i, Y_i\} \subset i \epsilon [0, t_n ]\}$ Mini-batch of training samples for continual learning }
\KwOut{Updated classifier model ($f_m$ )} 

\nl At $T_i \forall i \epsilon  [0, 1, 2, ... , N]: $

\nl \hskip0.25em $F_\theta, F_h \leftarrow DQ \{f_\theta, f_h\}$ 

{\fontfamily{qcr} \selectfont //training loss} \\
\nl \hskip0.25em $ l \leftarrow L_{CE}(F_h(F_\theta(X_i \epsilon M_{CL})))$ 

{\fontfamily{qcr} \selectfont //backpropagation} \\
\nl \hskip0.25em $ \frac{\delta l}{\delta \theta} \leftarrow BP(L_{CE}, Y_i \epsilon M_{CL})$ 

\nl \hskip0.25em $ w_m \leftarrow |F| \forall F_\theta, F_h $

{\fontfamily{qcr} \selectfont //model update} \\
\nl \hskip0.25em $ F^{'} \leftarrow f - \frac{\delta l}{\delta \theta} \odot w_m  \forall (F_\theta, F_h)  $   

{\fontfamily{qcr} \selectfont //quantize updated model} \\
\nl \hskip0.25em $\{f_\theta, f_h\} \leftarrow Q \{ F^{'} \forall (F_\theta,F_h)\}$ 

{\fontfamily{qcr} \selectfont //prototype update} \\
\nl \hskip0.25em $\{P_{class}\} \leftarrow (1/N)\sum_{i=1}^{N} f_\theta(X_i \epsilon M_{CL}) $ 

\caption{{\bf Algorithm for continual learning} \label{ContinualLearning}}

\end{algorithm}}

\begin{document}

\title{Domain-Incremental Continual Learning for Robust and Efficient Keyword Spotting in Resource Constrained Systems}

\author{
   Prakash Dhungana, 
   Sayed Ahmad Salehi   
   \\ University of Kentucky
}

\maketitle

\begin{abstract}

Keyword Spotting (KWS) systems with small footprint models deployed on edge devices face significant accuracy and robustness challenges due to domain shifts caused by varying noise and recording conditions. To address this, we propose a comprehensive framework for continual learning designed to adapt to new domains while maintaining computational efficiency. The proposed pipeline integrates a dual-input Convolutional Neural Network, utilizing both Mel Frequency Cepstral Coefficients (MFCC) and Mel-spectrogram features, supported by a multi-stage denoising process, involving discrete wavelet transform and spectral subtraction techniques, plus model and prototype update blocks. Unlike prior methods that restrict updates to specific layers, our approach updates the complete quantized model, made possible due to compact model architecture. A subset of input samples are selected during runtime using class prototypes and confidence-driven filtering, which are then pseudo-labeled and combined with rehearsal buffer for incremental model retraining. Experimental results on noisy test dataset demonstrate the framework's effectiveness, achieving 99.63\% accuracy on clean data and maintaining robust performance (exceeding 94\% accuracy) across diverse noisy environments, even at -10 dB Signal-to-Noise Ratio. The proposed framework work confirms that integrating efficient denoising with prototype-based continual learning enables KWS models to operate autonomously and robustly in resource-constrained, dynamic environments.

\end{abstract}


% Chapter:      Introduction
\section{Introduction}  \label{Introduction}

\IEEEPARstart{K}{eyword} Spotting (KWS) focuses on recognizing keywords from an audio input, often serving as an essential component to enable speech-based user interactions on smart devices. With the proliferation of voice assistants and intelligent control systems, KWS systems have been used on resource-constrained embedded systems and edge devices for real-time operation. Consequently, KWS models must prioritize low complexity, minimal memory footprint, and high energy efficiency to enable fast and real-time implementation.

Despite success in clean audio, the accuracy of KWS models severely degrades (up to 27\%) when deployed to the working environment that differs from the environment where training dataset was acquired \cite{efficient_rtkws_SpectroNet}. This is due to the discrepancies such as new speakers, accents, or background noise commonly known as domain shifts. Continual Learning (CL) addresses this critical challenge by continually updating the model to accommodate new domain data. It is essential to preserve previously acquired knowledge while continually updating the model to new data in order to achieve reliable performance; otherwise the model fails to retain the earlier acquired knowledge, which results in catastrophic forgetting. By allowing models to autonomously adapt to environmental and data shifts, CL facilitates robust operation without the need for repeated retraining from scratch. Research in this area is increasingly focusing on class-incremental, task-incremental, and domain-incremental CL. Class-incremental CL focuses on adding new classes of data that are encountered after initial training \cite{prototypeaug_sscil}, whereas task-incremental CL focuses on similar task with same number of classes but the classes within the task may vary \cite{LwF}. Domain-incremental CL focuses on the same classification task  but in a different domain context, such as background noise, input distributions, variations in speech tone \cite{ODDL}. This paper focuses on domain-incremental CL to address varying domains and varies with domain adaptation as it focuses on adapting to a single domain or a combination of few domains.

We propose an on-device CL framework for efficient KWS, introducing a robust framework for continual domain-incremental learning. Our approach integrates a quantized neural network model with CL strategies designed for embedded and edge use-cases. We utilize a combination of Wavelet Denoising and Spectral Denoising to extract reliable audio features. Crucially, our system uses a rehearsal memory buffer containing a subset of Mel Frequency Cepstral Coefficients (MFCC) and Log Mel-Spectrogram (LogMel) features from the initial training data for the adaptation and CL phase. This rehearsal buffer, along with its augmented versions (augmentation uses spectral feature map of one second recording from deployed environment), is combined with newly identified effective samples during runtime and assigned pseudo-labels based on classification confidence and prototype similarity to form the mini-batch for retraining. Contrary to some prior methods that only update the classifier layer or partial model architectures, our methodology updates the complete model incrementally, along with updating class prototypes during CL.

Our contributions advance the field of on-device CL by offering a comprehensive solution for domain-incremental learning: (1) We introduce a new CL framework for domain-incremental learning that updates the classifier model, leveraging pseudo-labeled effective samples as well as rehearsal buffer. (2) We integrate both wavelet transform and spectral denoising within the feature extraction pipeline to ensure robust feature maps for accurate classification and effective sample determination during runtime. (3) We demonstrate the robustness of using a dual feature input classifier model (MFCC + LogMel), which yields high accuracy (99.63\% on clean test data) and maintains high performance under noise. (4) We validate the performance of the CL framework in a wide range of noise levels (from $-10$ dB to $10$ dB SNR) and diverse environments, showing consistently high accuracy even in highly degraded conditions.

\section{Background and Motivation} \label{background_and_motivation}

\subsection{On device deep learning approaches}

Deploying deep learning-based KWS systems directly on edge devices requires models optimized for low latency, minimal memory footprint, and energy efficiency. Convolutional architectures that employ depthwise separable convolutions, such as MobileNet\cite{mobilenets}, significantly reduce computational complexity and parameters while maintaining robust accuracy. Similarly, Broadcasted Residual Networks (BC‑ResNet)\cite{BC-ResNet} enhance feature extraction by distributing residual connections over both time and frequency axes, yielding state-of-the-art accuracy(98\%) in speech-command benchmarks \cite{gscd} with reduced inference cost. Quantization methods—including post-training quantization (PTQ) and quantization-aware training (QAT) \cite{tflite_quantization} approaches—such as binary neural networks (e.g., 1D-BCNN \cite{1D-BCNN}) and integer-only schemes \cite{rtkws} further compress models while maintaining performance, making them ideal for embedded applications. Collectively, these approaches form the basis for high-performance, resource-efficient KWS suitable for embedded environments.

Earlier works that employs complex model architectures such as MobileNet\cite{mobilenets}, MCUNet \cite{mcunet, mcunetv2}, ODDA using DSCNN \cite{ODDA} have significantly higher parameter count and cannot be deployed for embedded environments with stringent memory requirements. In addition to model inference, to enable on device learning, an additional memory  and computation is required which increases latency for each samples being processed. Among existing works, Lin et. al. \cite{mcunetv3} has enabled on-device learning limited to certain specific layers and parameters using their proprietary methods that include Quantization Aware Scaling (QAS), sparse update, and Tiny Training Engine (TTE) in a single solution. ODDL \cite{ODDL} retrains the classifier layer of the DSCNN model architectures with features obtained from pre-labeled samples mixed with noise at 0 dB signal to noise ratio (SNR) and validates on two category of noise sources, namely \textit{seen} and \textit{unseen} during training.    

\subsection{Continual learning for domain adaptation}

Real-world deployment of KWS systems often involves encountering new speakers, accents, background noise, and various recording conditions. Conventional training fails to handle these scenarios effectively. Scenarios different from the conditions at which the training dataset is obtained are usually called domain shifts. CL addresses this challenge by continually updating the model to accommodate new domain data while preserving previously acquired knowledge during conventional training. Vu et al. \cite{efficient_cl_bnn} introduced a binary neural network for KWS and evaluated CL performance, achieving high accuracy with minimal memory overhead using the classifier layer, which is expanded to incorporate each new class encountered during class-incremental CL. Michieli et al. \cite{TAP-SLDA} presented TAP‑SLDA, which updates Gaussian prototypes incrementally for embedded devices without accessing stored samples, achieving up to 11$\%$ accuracy improvement when compared to Gaussian prototype without any updates. For domain adaptation application, Cioflan et al.  \cite{ODDA} demonstrated that on-device adaptation using merely 100 labeled samples and less than 10 kB read-write memory for intermediate activations can recapture up to 14$\%$ of the degraded performance in noisy environments. Unsupervised pseudo-labeling techniques have been successfully leveraged to personalize KWS systems \cite {personalized_KWS}. These literature highlight the importance of integrating efficient on-device inference with CL for domain adaptation. By enabling models to continually retrain themselves in response to environmental and data shifts, these models can operate robustly and autonomously in privacy-sensitive, resource-limited contexts without repeated retraining from scratch.

Earlier work that emphasizes CL has primarily focused on task-incremental and class-incremental CL applications [cite a couple of review papers]. Few works have been published that perform domain adaptation \cite{ODDL, mcunetv3}. The goal of our paper is to address domain-incremental learning that focus on varying domains in addition to domain adaptation. The learning mechanism utilized in \cite{ODDL} requires parallel processing capabilities and can be implemented using complex hardware such as GAP9 processor. The noise sources are divided into seen (used for training) and unseen noises (used for evaluation). However, in our work, we assume that the noises occur after deployment (unseen noises) and any audio deformations encounterd should be adapted during runtime. Lin at. al. \cite{mcunetv3} have implemented their proprietary solution using simulated embedded environment using highly capable GPUs usually utilized for conventional training and optimized for memory requirements to fit embedded applications. Contrary to ODDL \cite{ODDL}, Lin et.al. \cite{mcunetv3} updates multiple layers of the utilized model in addition to the classifier layer using sparse update. In our work, we propose a new framework for CL designed for domain adaptation that updates the complete model (contrary to both \cite{ODDL, mcunetv3}) and validate the performance at multiple noise levels (contrary to 0 dB SNR used in \cite{ODDL}) using pre-labeled rehearsal samples and runtime effective samples (described in section \hyperref[effectivesamples]{IIIB}).      

\section{Proposed Methodology} \label{Proposed_Methodology}

The overview of our proposed CL framework for domain-incremental learning is shown in \hyperref[figure_overviewCL]{Fig 1}. This framework integrates quantized neural network model obtained from initial training (classifier model - see Section IIIA4 and IIIA5) with CL strategies specifically designed for resource-constrained applications. This framework is implemented in two stages, namely, determination of effective samples and CL for repetitive domain adaptations at fixed intervals. The first stage, determines if a run-time sample (i.e., input sample during run-time) should be flagged as an effective sample or not. If the sample is flagged as effective sample, it is saved in memory to be used during the second stage. During the second stage (CL phase), effective samples are combined with existing rehearsal buffers (including augmented versions) to form a mini-batch of training samples. Rehearsal buffers contains labeled MFCC and LogMel features of a subset of initial training dataset. The mini-batch is used to retrain the classifier model (see Section IIIC) during model updates. After each instance of model retraining during the CL phase, the class prototypes, the thresholds for the prototype distances, and the quantized classifier model are updated using the retrained model. %The utilized approaches are described in their respective sections.

\figureoverviewCL

This section begins with the description of the KWS pipeline used for our proposed CL framework. KWS pipeline provides details about audio processing, feature extraction, spectral denoising and the classifer model used in the CL framework. After KWS pipeline, we describe how effective samples are identified using the input sample encountered are runtime. Finally, we provide an explanation of the CL phase that encapsulates the KWS pipeline and effective samples to address domain-incremental learning after deployment.  

\subsection{KWS Pipeline}

Our design is based on compact CNN model architecture similar to \cite{rtkws}, \cite{horakws} and is composed of four distinct components, namely wavelet denoising, feature extraction (FE), spectral denoising and classifier model. Wavelet denoising (Section IIIA1) is used to denoise the audio using raw waveforms. Feature extraction (Section IIIA2) generates a feature map from denoised audio to be utilized as input for spectral denoising. Spectral denoising (Section IIIA3) performs the final denoising operation directly on MFCC and LogMel using spectral and temporal information present in the feature map. Classifier model architecture subsections (Section IIIA4 and IIIA5) describe the model architecture, quantization and deployment used for generating the KWS models.

\subsubsection{Wavelet Denoising}
The wave samples acquired from ADC are processed with Haar wavelet transform \cite{haarwavelet} applied over sliding frames. The input signal is segmented into non-overlapping frames of length $N$ (1024 samples). For each frame, single level wavelet decomposition is performed. The transform separates each frame into approximation and detail coefficients, where the latter primarily capture high-frequency noise. Denoising is performed using VisuShrink with universal thresholding \cite{universalthresholding}. VisuShrink is one of the popular denoising method that removes noise by applying a single global “universal” threshold to restrict the wavelet coefficients, prioritizing smooth reconstructions over preserving fine details.

Noise variance is estimated using the median absolute deviation (MAD) of the first-level detail coefficients shown in Equation (1). Then a threshold $\tau$ is calculated using MAD and window length using Equation (2). Coefficients below $\tau$ are suppressed via soft thresholding to attenuate noise while preserving higher detail coefficients. The denoised frame is reconstructed through the inverse Haar wavelet transform and combined with adjacent frames. The output of the wavelet denoising is quantized to range $[-128, 127]$. This method effectively removes nonstationary noise while preserving the temporal and spectral integrity of the audio signal. The output denoised waveform is then subsequently used for feature extraction using CMSIS library which is discussed in the next section.

\begin{equation}
    MAD = \frac{median(|d - median(d)|)}{0.6745}
\end{equation}

\begin{equation}
    \tau = MAD * \sqrt{2*log(N)} 
\end{equation}

where d represents the detail coefficients obtained from the Harr wavelet, \textit{MAD} is the median absolute deviation, $\tau$ is the universal shrinkage coefficient used as soft threshold for wavelet denoising, and N is the window length. 

\subsubsection{Feature Extraction}

The feature extraction part transforms the denoised audio waveform into a 2-dimensional (2D) spectral feature map. Mel-Spectrogram and Mel Frequency Cepstral Coefficients (MFCC) are widely used in KWS \cite{CNN_Interspeech_Sainath, DSCNN_IEEE_Miah, hello_edge, kws_using_gcn, mdpi_purdue_thesis, edgernn, deep_residual_learning_kws}. However, these algorithms include floating point operations and require significant memory, inefficient for low-cost and real-time implementation on embedded systems. To avoid floating point operations, save memory and latency, we have utilized CMSIS-DSP \cite{ARM_CMSISDSP} library, which supports integer-based MFCC computation. Additionally, Mel-Spectrogram is an intermediate feature map within MFCC computation pipeline and is used as input feature map in conjunction with MFCC to improve performance of the classifier model without incurring additional compute and latency during feature extraction.

\figuredenoising

\subsubsection{Spectral Denoising}

Spectral denoising begins with normalizing the input feature map and then leveraging both temporal and spectral masks from the feature map to suppress noise. Each feature map is first scaled to the range $[0,1]$, ensuring invariance to amplitudes and providing a stable basis for further processing. This guaranties consistent performance on diverse inputs with varying noise information. Following normalization, mean subtraction was performed along two axes: the temporal axis (removing frame-level averages) and the spectral axis (removing frequency-wise averages), as shown in Equations 3. These operations address any signal variations caused by background noise on their respective axes, which tends to be evenly distributed. A masking step then selectively retains only components above the mean, effectively preserving prominent signal structures (e.g., speech harmonics or transients) and suppressing weaker noise-like elements as shown in Equations 4. The temporal (second column) and spectral (third column) masks obtained for a random sample in one of the noisy environment used during the evaluation are shown in \hyperref[figure_denoising]{Fig 2}. 

\begin{equation} \label{equation:3}
\begin{split}
    x_{t} = x_{n} - \mu_t (x_{n}) \\
    x_{s} = x_{n} - \mu_f (x_{n})
\end{split}
\end{equation}
\begin{equation} \label{equation:4}
\begin{split}
    M_{t} = 1 [x_{t} > \mu_t (x_{t})] \\
    M_{s} = 1 [x_{s} > \mu_f (x_{s})]
\end{split}
\end{equation}
\begin{equation} \label{equation:5}
    x_{d} = (1 - \alpha) ( \alpha x_{s} M_{s} + (1 - \alpha) x_{t} M_{t} ) + \alpha x_{n}
\end{equation}

where $x_{n}$ is the normalized feature map (MFCC or Mel Spectrogram), $\mu_t (x_{n})$ is the mean across the temporal axis of $x_{n}$ , $\mu_f (x_{n})$ is the mean across the frequency axis of $x_{n}$, $ x_{t}$ and $x_{s}$ are the intermediate feature obtained by subtracting $\mu_t (x_{n})$ and $\mu_f (x_{n})$, respectively,  $M_{t}$ and $M_{s}$ are masks that estimates speech and noise presence respectively, $\alpha$ is an attenuation parameter to control the degree of denoising and $x_{d}$ is the final denoised output feature.

Finally, to obtain the denoised feature map (last column in \hyperref[figure_denoising]{Fig 3}), we perform a recombination of the mean subtracted features (temporal and spectral) with the input feature map using the temporal and spectral masks and the attenuation coefficient $\alpha$. The attenuation parameter $\alpha$ controls the trade-off between aggressive noise suppression and signal preservation: smaller values emphasize denoising, while larger values retain more of the original data. This denoising method integrates ideas from spectral subtraction, masking, and normalization into a lightweight and dynamic method suitable for time–frequency representations of noisy audio.

\subsubsection{Single feature map model architecture}

Convolutional Neural Networks (CNNs) stand out as the predominant model for image classification because of their superior accuracy compared to alternative models. This superiority is credited to the efficacy of 2-dimensional convolution (Conv2D) layers, which form the fundamental building blocks of CNN architectures. Taking advantage of the fact that the feature representation extracted from audio signals inherently resembles a 2D image, the integration of Conv2D layers proves exceptionally effective in audio classification tasks. Previous research, as cited in \cite{CNN_Interspeech_Sainath}, has underscored the remarkable accuracy achievable in audio classification through the use of Conv2D layers. In contrast, Fully Connected (FC) layers pose challenges such as excessive model parameters and the oversight of spatial and local features. Conv2D mitigates these limitations of FC layers by employing streamlined kernel matrices for convolution operations on input feature maps, thus enhancing performance and efficiency.  

\figuresinglefeaturemodel

Taking advantage of the benefits offered by Conv2D layers over FC layers, the proposed model architecture is structured to incorporate three consecutive Conv2D layers, followed by a Flattening layer and a final FC layer, as illustrated in \hyperref[figure_singlemodelarchitecture]{Fig 3}. Each Conv2D layer employs a kernel size of 5x5 and utilizes a rectified linear unit (ReLU) activation function. The first layer processes a 20x16 MFCC output from feature extraction, which yields a 16x12 feature map. Similarly, the second and third Conv2D layers generate intermediate feature maps of dimensions 12x8 and 8x4, respectively as seen in \hyperref[figure_singlemodelarchitecture]{Fig 3}. Furthermore, as experimentation reveals minimal impact of zero padding on accuracy, the Conv2D layer operations are executed without zero padding to reduce memory footprint and computational demands for subsequent layers. Prior to the final FC layer, a Flattening layer is utilized, with the FC layer containing 160 parameters. Initially, the model is trained using 32-bit floating point precision followed by QAT. The model obtained after QAT is then used as a starting point for CL.  

The intermediate feature after the third Conv2D is used as the latent representation that will be used to compare with the class prototype. This location for latent representation was chosen considering the memory and computational requirement to determine effective samples during runtime. The size of the latent representation at this location is 160 bytes compared to 960 and 192 bytes after the first and second Conv2D layers, respectively. Additionally, to compare the latent representation with the class prototype, 160 and 320 arithmetic operations is required using MAE and MSE methods, respectively. Details on class prototypes and how effective samples are determined are provided in {Section IIIB}.

\figuredualfeaturemodel

\subsubsection {Dual feature map model architecture}
The dual-input CNN is designed to leverage the strengths of two distinct feature representations. The architecture begins with two parallel input streams that accept MFCC and Log Mel-spectrogram features, as shown in \hyperref[figure_dualmodelarchitecture]{Fig 4}. Each stream independently processes its input through a series of three sequential 2D convolutional layers (similar to single feature model architecture), allowing the network to learn specialized patterns from both the cepstral and spectro-temporal domains. Following the convolutional layers, the resulting two-dimensional feature maps from each path are transformed into one-dimensional vectors by their respective flattening layers. These vectors are then fused into a single vector, through a Concatenation layer, effectively combining the learned features from both path. The concatenation layer is chosen as the latent representation for the class prototype in order to incorporate information from both paths and because of the reasoning provided in the single input model architecture. The size of the latent representation and class prototypes for this model architecture is twice that of the single feature model. Finally, a fully connected Dense layer performs the ultimate classification based on this unified feature set, enabling a robust decision-making process that benefits from the two types of input features.

\subsection{Effective samples Determination} \label{effectivesamples}
After deployment, the input samples encountered during runtime undergoes quantized inference for classification. After each quantized inference, if the classification probability is above a certain threshold, the input is assigned the label of the predicted class. Finally, the latent representation of the input is compared with the prototype of its assigned class. If the distance between the prototype and the latent representation is within the acceptable threshold, then the input sample is determined as an effective sample and is saved in memory to be used during CL phase. Further details regarding determination of effective samples are provided in the following subsections.

\subsubsection{Class Prototypes}
Prototype-based classification offers an efficient and interpretable method for classification and posterior handling of a neural network model by representing each class through a prototype \cite{Protonets_FSL}. Typically, prototypes are computed as the mean of latent representations derived from the feedforward operation of a neural network model with training samples or projected from the intermediate features into the same latent space (commonly known as latent representation). These latent representations reside in a feature space optimized to capture the discriminative properties of the input feature map of each class. Using a latent representation as a class prototype mitigates the task of training auxiliary networks to obtain class prototypes and also reduces the computation and memory requirements \cite{self_train}. \iffalse If an auxiliary network is used, the weights associated with the network increase the memory requirement. In addition to that, feature projection from the intermediate features to the latent representation increases compute requirements. \fi During the inference operation, test samples are embedded in the same latent space and assigned to the class whose distance from the prototype is closest under a chosen distance metric, such as MAE, MSE or RMSE. We have used MAE for calculating distance in our experiments. Theoretically, this approach aligns with the principles of metric learning, which posit that semantically similar instances are clustered. Neural networks, through supervised training, learn transformations that map input to separable regions in latent space. In this context, class prototypes approximate the centers of these regions, enabling simple but effective nearest-prototype classification. 

Additionally, using class prototypes facilitates posterior handling without requiring full model retraining, which is particularly advantageous in continuous or a few-shot learning settings. It supports memory-efficient learning by storing only representative embeddings rather than full datasets, and enables rapid adaptation by updating prototypes as data distributions evolve. Thus, prototype-based classification combines the representational power of feature embeddings with the simplicity of distance-based decision rules, yielding a scalable and theoretically grounded solution for dynamic and resource-constrained learning scenarios. Furthermore, in our implementation, class prototypes are used as the method to assign pseudo labels to the audio samples based on which the distance from the respective class prototype is calculated. 

\figureeffectivesamples

A subset of MFCC and Mel Spectrogram computed using the training data from Google Speech Commands Dataset (GSCD) is used as a rehearsal buffer. The rehearsal buffer is saved in memory, which is used to generate augmented samples using the MFCC and Mel-Spectrogram of noise recorded from the deployed environment. The augmented sample only undergoes spectral denoising as rehearsal buffers are already on feature domain. The primary goal of using feature maps as rehearsal samples is to reduce memory overhead and improve computational efficiency. During CL, samples identified as effective during runtime as well as rehearsal buffer and augmented rehearsal buffer are included into the mini-batch, as illustrated by the dotted path in \hyperref[figure_effectivesamples]{Fig 5}. \textbf{Latent representations obtained using samples from mini-batch and feedforward operation of the quantized model are used to compute class-wise artifacts that includes: class prototype, mean distance and deviation observed within the mini-batch}. A single sample is passed through the quantized inference and retrieves two outputs: an intermediate latent representation and the final prediction from the classifier model. For each corresponding class within the mini-batch, the latent vectors are used to compute the class prototypes. Each prototype is obtained by taking the mean across all latent representations of samples within the mini-batch for its respective class and has the same size as the latent representation. After obtaining class prototypes, to evaluate how well each sample aligns with its class prototype, we calculated the MAE between the latent representation of the sample and its corresponding prototype. These distances indicate the distribution of latent representation of the class with it's prototype. We then compute the mean and standard deviation of the distances calculated. The results (class prototypes, mean distance, and standard deviation) are utilized later when running quantized inference of the runtime samples to determine effective samples, which is described in the next Section. 

\subsubsection{Effective Samples} \label{effectivesamples]}

Effective samples are identified using a combination of confidence-driven filtering (reflecting classification confidence) and prototype-based similarity (reflecting consistency with class prototypes). Denoised feature maps from sampled audio is fed to quantized inference to obtain latent representation as described earlier, and its distance (MAE) is calculated with respect to the class prototype. This distance acts as a measure of the degree to which the sample aligns with the prototype of its predicted class. Only samples with high classification confidence and a low distance from their class prototype are selected as effective samples. We consider a single threshold for distance. We use ($\mu + n\sigma$) as threshold for prototype distances as shown in \hyperref[figure_effectivesamples]{Fig 5} based on empirical evaluation. The intuition behind thresholding is that we only select samples whose distance from class prototype is within acceptable range and outliers are neglected from being flagged as an effective sample. With the threshold for distance from it's prototype, we are limiting the samples that provides new information to the model and the distance beyond which ambiguous information for class prototypes might be introduced. As seen in \hyperref[figure_effectivesamples]{Fig 5}, samples that are present in the green region (yellow dots) are selected, while samples that are outside the green region (red dots) are not selected. This strategy ensures that selected samples are both confidently predicted and representative of their class distribution supporting more stable learning in CL scenario for dynamic environments.

\algorithmone

During run-time, with consideration for embedded systems, it is necessary to perform data sampling and model inference in real time. During the model inference, our proposed algorithm for the determination of the effective sample is also run along with the quantized model inference. Quantized inference is performed in INT8 precision. During deployment, class artifacts are calculated using a mini-batch formed from rehearsal and augmented rehearsal buffer. During run-time, the augmented rehearsal buffer is updated with the new information from the environment. Class artifacts are provided to \hyperref[algorithm_one]{Algorithm I} as inputs.

\hyperref[algorithm_one]{Algorithm I} provides details on determining whether a sample is effective by evaluating its confidence score and its similarity to the class prototype using distance metric in the latent space. The model inference operation is performed in INT8 due to which the confidence of predicted output is also calculated in integer (Line 3 - Algorithm I). The confidence score threshold to be flagged as effective sample is 85\% (emperical evaluation, discussed in Section V). During quantized inference, the predicted output of the sample is obtained and compared with the confidence score. If the predicted output have confidence score higher than the threshold, then only we proceed with calculating the distance from the class prototype. Line 6 (Algorithm I) shows the computation of MAE distance from the class prototype. By comparing the confidence score first, we aim to optimize any redundant computation of distance calculations that might not be necessary if the confidence score is not met. Otherwise, the distance of the latent representation of the runtime sample with the class prototype of predicted class is computed. If the distance is within the acceptable range, the sample is then identified as an effective sample and labeled as the predicted class. 

\algorithmtwo
    
\subsection{Continual Learning}
CL is crucial for sustaining robust performance in dynamic, real-world environments that subjected to domain shifts. \hyperref[algorithm_two]{Algorithm II} outlines a systematic approach for enabling CL, specifically designed to update a classifier model incrementally using mini-batches of training samples formed using rehearsal buffer, augmented rehearsal buffer and effective samples (that are pseudo-labeled). The primary objective is to efficiently update the model while addressing issues such as catastrophic forgetting and concept drift. The algorithm begins by processing mini-batches $M_{CL}$ containing training samples $\{X_i, Y_i\}$. Each training iteration involves quantization-aware operations indicated by the function $DQ\{f_m\}$, extracting a dequantized representation $F_{m_{fp}}$ of the model $f_m$. Training loss $l$ is calculated via the cross-entropy loss function $L_{CE}$, which is used in the subsequent backpropagation step for gradient calculations.

After obtaining the gradients, the model is updated by modifying the weights with the same learning rate (0.001) utilized during conventional training, thus controlling the magnitude of updates for stable re-training. 

Finally, quantization ($Q$) is applied to update the model's parameters into an INT8 format, aligning with practical deployment constraints for complete integer quantized inference. Simultaneously, class prototypes $P_{class}$ are updated to encapsulate new learning insights effectively using the training mini-batch along with mean and deviation values of the distances from respective class prototypes. Model quantization coupled with prototype update ensures model compactness and efficient inference capability, essential for real-time deployment on resource-constrained edge devices. Overall, this algorithm encapsulates a robust CL procedure capable of dynamically adapting to new scenarios and the retention of previously acquired knowledge.
 
\section{Results} \label{Results}
\subsection{Dataset and Experimental setup}
Google speech commands dataset (GSCD v2) was used as training dataset. The dataset consists of over 105,000 one-second or less audio files sampled at 16 kHz of 35 different keywords, spoken by 2,618 different speakers. Keywords ``Yes" and ``No" are used from the dataset for binary classification. Testing dataset from GSCD v2 is used to evaluate the performance of classifier models. 

Additionally, for simulating the deployed environment, DEMAND dataset is used as the noise sources. DEMAND dataset contains noise sources from six categories of environments - Domestic, Nature, Office, Street, Public and Transportation. Each categories has three different recording environments. The noise samples are 300 seconds in duration and recorded at 48 kHz and resampled at 16 kHz as well. We have utilized four of the five categories selecting single recording from each category. 

For evaluating the proposed framework for domain adaptation, evaluation samples are generated using audio mixing at multiple SNRs from -10 dB to 10 dB with 5 dB increments to simulate the actual runtime environment. Each test sample from the GSCD v2 dataset is mixed with single second noise data from the DEMAND dataset. Thus, there are 169,695 evaluation samples in total for each noisy environment. During evaluation, the classifier models are adapted to the environment using spectral domain feature mixing using the rehearsal buffer and random noise samples from the 300 seconds period. After initial adaptation, the classifier models runs in inference mode. During inference operations, the effective samples are determined (as discussed in section IIIB) and saved in memory for next adaptation phase of the CL. The period between consecutive model retraining during the CL phase are kept at 1024 samples which is composed of random 8 sec of noise audio from the 300 sec mixed with 64 samples each of ``Yes" and ``No" samples respectively. The average accuracies are reported after 25 iterations of model retraining have been completed.     

\modelaccuracytable

\finalresults

\componentsaccuracies

\begin{table*}[]
\centering
\captionsetup{justification=centering}
\caption{Comparison of existing on-device learning framework with our framework.}
\label{table:4}
\begin{tabular}{|cccccccc|}
\hline
\multicolumn{8}{|c|}{Measurements   per sample} \\ \hline
\multicolumn{1}{|c|}{Architecture} &
  \multicolumn{1}{c|}{Dataset} &
  \multicolumn{1}{c|}{Hardware} &
  \multicolumn{1}{c|}{Parameters} &
  \multicolumn{1}{c|}{FLOPS} &
  \multicolumn{1}{c|}{Latency (ms)} &
  \multicolumn{1}{c|}{Energy (µJ)} &
  Accuracy (\%) \\ \hline
\multicolumn{1}{|c|}{MCUNetv3} &
  \multicolumn{1}{c|}{VWW} &
  \multicolumn{1}{c|}{STM32F746 (M7)} &
  \multicolumn{1}{c|}{480 k} &
  \multicolumn{1}{c|}{46 M} &
  \multicolumn{1}{c|}{546} &
  \multicolumn{1}{c|}{$6899^*$} &
  89.3 \\ \hline
\multicolumn{1}{|c|}{ODDL DSCNN S} &
  \multicolumn{1}{c|}{GSCD} &
  \multicolumn{1}{c|}{GAP9} &
  \multicolumn{1}{c|}{23.7 k} &
  \multicolumn{1}{c|}{2.95 M} &
  \multicolumn{1}{c|}{6.74} &
  \multicolumn{1}{c|}{384} &
  90.68 \\ \hline
\multicolumn{1}{|c|}{ODDL DSCNN M} &
  \multicolumn{1}{c|}{GSCD} &
  \multicolumn{1}{c|}{GAP9} &
  \multicolumn{1}{c|}{138.1 k} &
  \multicolumn{1}{c|}{17.2 M} &
  \multicolumn{1}{c|}{16.34} &
  \multicolumn{1}{c|}{974} &
  92.64 \\ \hline
\multicolumn{1}{|c|}{ODDL DSCNN L} &
  \multicolumn{1}{c|}{GSCD} &
  \multicolumn{1}{c|}{GAP9} &
  \multicolumn{1}{c|}{416.7 k} &
  \multicolumn{1}{c|}{51.1 M} &
  \multicolumn{1}{c|}{32.95} &
  \multicolumn{1}{c|}{2028} &
  93.47 \\ \hline
\multicolumn{1}{|c|}{\multirow{2}{*}{Ours}} &
  \multicolumn{1}{c|}{GSCD} &
  \multicolumn{1}{c|}{TM4C123GXL (M4)} &
  \multicolumn{1}{c|}{1.64 k} &
  \multicolumn{1}{c|}{0.89 M} &
  \multicolumn{1}{c|}{$95.11^{**}$} &
  \multicolumn{1}{c|}{$200.87^{**}$} &
  94.31 \\ \cline{2-8} 
\multicolumn{1}{|c|}{} &
  \multicolumn{1}{c|}{GSCD} &
  \multicolumn{1}{c|}{Optimal M4} &
  \multicolumn{1}{c|}{1.64 k} &
  \multicolumn{1}{c|}{0.89 M} &
  \multicolumn{1}{c|}{$95.11^{**}$} &
  \multicolumn{1}{c|}{$64.45^{**}$} &
  94.31 \\ \hline
\end{tabular}
\begin{tablenotes}
\item {* : Estimated using ARM M7 Processor Datasheet}
\item {** : Estimated using ARM M4 Processor Datasheet}
\end{tablenotes}
\end{table*}

\subsection{Performance on GSCD}

\hyperref[table:1]{Table I} show the accuracies of the single input and dual input classifier models on test data from GSCD v2. The accuracy of the dual input classifier model is higher than the single input classifier model. The higher accuracy achieved by the model can be attributed to the learning enabled by the different spectro-temporal features provided by MFCC and Mel Spectrogram. From this point forward, we use dual input classifier model for CL and evaluation purposes.     

\subsection{Performance on Noisy Dataset}

The results of domain adaptation for multiple noise environments are provided in \hyperref[table:2]{Table II}. \hyperref[table:2]{Table II} summarizes the performance of the CL on four distinct environments, namely DWASHING, NFIELD, OOFFICE, and TCAR. The table reports accuracy across five SNR levels (-10 dB to 10 dB), allowing assessment of the degree of noise robustness after continual updates. As the quality of the audio degrades, the performance of the classifier model also degrades. However, the degree of the reduction in performance is minimal because of CL even for higher level of noise (i.e. -10 dB).   

Across all environments, CL leads to consistently high performance even under higher noise levels. The adapted model on DWASHING environment achieves strong accuracy at -10 dB (93.84\%) and continues improving with increasing SNR, indicating that this environment achieves a stable adaptation. The NFIELD and OOFFICE models show slightly lower accuracy at the lowest SNR levels but remain highly competitive, with both exceeding 94\% accuracy by 0 dB and maintaining stable performance at higher SNRs. These results reflect the effectiveness of the CL framework in quickly aligning the model to the runtime environments. The highest overall accuracies are observed for the TCAR environment, which achieves 94.56\% at -10 dB and peaks at 95.28\% at 0 dB. This suggests that TCAR environment also achieves stable adaptation. Overall, Table II demonstrates that even minimal CL (only 25 updates) significantly boosts noise robustness across diverse environments, enabling the classifier model to retain strong accuracy from low-SNR to clean conditions.

\figureablationalpha

\subsection{A closer look}
\hyperref[table:3]{Table III} reports the classification accuracies obtained on the OOFFICE runtime dataset across five SNR levels ($-10\text{ dB}$ to $10\text{ dB}$) and under different configurations of the proposed framework. The table systematically evaluates three factors: (i) model retraining during CL, (ii) the use of wavelet-based denoising, and (iii) the use of spectral denoising. Each row corresponds to a unique combination of these components, allowing a systematic assessment of their individual and combined contributions to noise-robust performance.

The first two rows represent performance without retraining, isolating the effect of denoising. When only spectral denoising is applied (Row 1), the model achieves competitive accuracy even at $-10\text{ dB}$, indicating that spectral denoising effectively diminishes noise information from the feature map. The addition of wavelet denoising along with spectral denoising (Row 2) slightly reduces performance at lower SNRs (e.g., $86.38\%$ at $-10\text{ dB}$), suggesting that this combination may introduce distortions when the signal is heavily corrupted. But the distortions introduced by wavelet denoising can be mitigated when retraining is enabled during CL (Row 4) which signifies that the distortions were from noise source rather than audio features. Rows 3 and 4 introduce model retraining, which consistently raises accuracy in the low-SNR conditions. With retraining and spectral denoising (Row 3), accuracy reaches $92.88\%$ at $-10\text{ dB}$—an improvement of more than 4\% when compared to the best configuration without retraining. When both denoising methods and retraining are combined (Row 4), the performance remains strong at very low SNRs and stable across higher SNR levels. Even though wavelet denoising in combination with CL offers marginal benefits at higher SNRs, in highly degraded conditions (-10 dB) it provides the best performance when compared to other results (Column 1) and is essential for robust performance.

\figureablationthprob
\figureablationthdist

\subsection{Analysis for Embedded systems}

\hyperref[table:4]{Table IV} compares the on-device learning frameworks designed for embedded systems, focusing on the efficiency metrics essential for TinyML deployment. This comparison showcases the trade-offs between model complexity (FLOPs and Parameters) and embedded system constraints (Latency and Energy) to achieve optimal performance (Accuracy) for specific tasks (VWW, KWS) across diverse low-power embedded platforms. The MCUNetv3 model, implemented on the STM32F746 (Cortex M7), is optimized for on-device training under tight memory budgets (e.g., 256 kB SRAM). Although achieving 89.3\% accuracy in the Visual Wake Words (VWW) dataset, its use of resources reflects a high computational load during training, with 46 M FLOPs and an estimated energy consumption of 6899 $\mu$J and latency of 546 ms per sample. In contrast, the ODDL is implemented on the multi-core GAP9 processor, targets On-Device Domain Adaptation (ODDA) for noise robustness using the rehearsal buffer. ODDL framework use Depthwise Separable CNNs (DS-CNN), demonstrating clear scalability: the largest model (L) uses 416.7 k parameters and 51.1 M FLOPS to achieve 93.47\% accuracy, while the smallest (S) achieves 90.68\% accuracy with only 23.7 k parameters.

In our CL framework, we use efficient CNN architecture optimized for KWS, which stands out due to its minimal footprint and exceptional accuracy on the GSCD. It utilizes only 1.64 k of parameters and requires the lowest computational complexity at 0.89 M FLOPS. This compact design enables deployment on the resource-constrained TM4C123GXL(ARM M4) microcontroller. The bottom two rows in Table IV highlight the critical impact of software/hardware co-optimization for embedded systems. Although both maintain the same high accuracy (94.31\%) and computational complexity, utilizing optimized M4 hardware results in significant gains for energy consumption, which is reduced from 200.87 $\mu$J to 64.45 $\mu$J per sample. The balance of high accuracy and extremely low complexity led to higher efficiency over its counterparts, as it achieves a comparable accuracy with 98.81\% fewer parameters than highly complex networks (DSCNN-S). 

\section{Ablation Studies} \label{Discussions}

\subsection{Denoising sensitivity analysis}

To evaluate the effectiveness and the impact of the  attenuation parameter ($\alpha$), we performed a sweep experiment for different values of $\alpha$. This experiment studies how the degree of spectro-temporal suppression relative to input affects the downstream classification accuracy across multiple environments. The results show that performance is generally stable for $\alpha \epsilon [0.4, 0.9]$, with only minor fluctuations across all SNRs. High-SNR conditions (0 to 10 dB) consistently maintain accuracies above 94–96\% for all environments, indicating that the denoising algorithm does not distort clean or mildly noisy signals even at more aggressive settings. In contrast, the lowest SNR condition (–10 dB) exhibits greater sensitivity, particularly in NFIELD and OOFFICE, where the accuracy occasionally dips when $\alpha$ becomes too dominant or too weak; however, the trend does not show a sharp optimum, suggesting robustness to parameter selection. Overall, this experiment demonstrates that the proposed denoising algorithm is highly stable, with no critical degradation throughout the operating range.

\subsection{Continual Learning – Confidence-based thresholding}

The probability-threshold ablation further evaluates the thresholding of runtime samples using posterior-probability confidence rather than the distance of latent representation from its class prototype. Similar to the sensitivity analysis of attenuation parameter, the probability-threshold curves remain nearly flat across all environments and SNRs. Performance changes are typically <0.5\%, showing that the classifier’s predictive confidence is well-calibrated and largely unaffected by moderate shifts in thresholding. Notably, even at the lowest SNR (–10 dB), accuracy remains stable as the threshold shifts from 0.70 to 0.85, indicating that the model retains reliable confidence separation between correct and incorrect predictions. The consistency across environments demonstrates that the continual-learning mechanism is insensitive to this hyperparameter and can operate effectively on various deployed environments.

\subsection{Continual Learning – Distance-based thresholding}

This study assesses how the thresholding using distance from class prototype during CL responds to variations in the threshold used for accepting or rejecting runtime samples using their latent representations. Across all environments, the results reveal remarkable stability: accuracy varies by less than 1–1.5\% across the entire sweep, indicating that the model’s latent representation decision boundary is naturally well-separated. High-SNR (0 to 10 dB) settings again show the highest and most stable accuracy, while low SNR (–10 dB) experiences slightly more variance, especially in NFIELD. Importantly, none of the noisy environments exhibit a sharply defined optimal threshold. Performance of CL plateaus across the entire range [1.7,2.4] of thresholds experimented. This implies that the distance-based CL is robust to hyperparameter choice and does not rely on precise tuning to maintain high accuracy, even under strong noise corruption.

\section{Related Work} \label{Related_Work}

\subsection{Domain adaptation}
Domain adaptation (DA) methodologies aim to improve model robustness when encountering discrepancies between training (source) and deployment (target) domains. Existing approaches generally fall into supervised, unsupervised, or semi-supervised paradigms. Unsupervised domain adaptation methods have drawn significant attention due to their ability to leverage unlabeled target domain data effectively. Notably, adversarial training frameworks such as Domain Adversarial Neural Networks (DANN) have demonstrated the capacity to align feature distributions between domains by employing a domain classifier in an adversarial training scheme \cite{domainadaptationbyBackprop}. Variants like conditional adversarial domain adaptation further refine this strategy by aligning feature distributions conditioned on the class labels, enhancing class-discriminative adaptation \cite{conditional_ada}. Recently, domain adaptation for keyword spotting specifically targeted lightweight networks, demonstrating effective performance recovery under constrained resource environments \cite{ODDL}. These studies underline the critical role of domain adaptation in ensuring practical utility and reliability of edge-based keyword spotting models in diverse real-world settings.

\subsection{Rehearsal based continual learning}
Rehearsal-based CL approaches maintain past knowledge by retaining representative samples or embeddings, typically stored in a rehearsal memory buffer. Incrementally retraining models on this buffer alongside new data mitigates catastrophic forgetting, a prevalent issue in neural networks. Prior works such as Rebuffi et al. have introduced incremental rehearsal-based methods (iCaRL) that combine representation learning with nearest-class-mean classification, effectively preserving previously acquired knowledge while accommodating new classes \cite{iCaRL}. Lopez-Paz and Ranzato proposed Gradient Episodic Memory (GEM), which uses memory-stored samples to constrain gradient updates, ensuring stable learning without degrading prior performance \cite{GEM}. Recent adaptations for embedded devices have sought to reduce memory footprint further by rehearsal of compressed embeddings or carefully selected prototypes, significantly improving efficiency while preserving CL capability \cite{A-GEM}. These methods demonstrate the practical feasibility of rehearsal-based strategies for on-device CL scenarios. However, these methodologies have been used in class and task incremental CL. We aim to use rehearsal based CL for domain-incremental learning. 

\subsection{Prototype-based Semi-supervised Continual Learning}
Prototype-based CL approaches leverage latent-space representations to create compact class prototypes, facilitating semi-supervised classification and efficient incremental learning. Typically, such methods employ neural network-extracted embeddings to form representative prototypes that serve as stable references during incremental adaptation. Michieli et.al. introduced TAP-SLDA, a semi-supervised CL algorithm that updates prototypes incrementally, maintaining classification accuracy without storing raw samples \cite{TAP-SLDA}. Similarly, Buzzega et.al. proposed Dark Experience Replay (DER), which leverages prototype-based knowledge distillation in rehearsal-free CL scenarios, enhancing performance without significant memory overhead \cite{DER}. Prototype-based semi-supervised strategies have further been combined with uncertainty estimation techniques, allowing effective selective updating of prototypes based on the quality and reliability of incoming samples, thereby improving robustness under label scarcity and distribution shifts \cite{Protonets_FSL, prototypeaug_sscil}. Collectively, these prototype-based frameworks highlight a promising pathway toward sustainable and efficient CL on resource-constrained edge devices. However, these techniques have not been utilized in conjunction with supervised CL. We aim to bridge this gap by encompassing knowledge from pseudo labeled effective samples during runtime with supervised retraining instances during CL.

\section{Conclusions} \label{Conclusions}
Domain adaptation including domain-incremental learning is expected to play a critical role in TinyML and federated learning by enabling scalable machine learning solutions on embedded platforms and devices. By mitigating performance degradation caused by domain shifts across heterogeneous sensors, operating conditions, and deployment environments, domain-incremental learning allows models to generalize without costly retraining or centralized data collection. This capability is particularly important in resource-constrained and privacy-preserving settings, where continuous on-device learning and robust cross-domain generalization are essential for sustainable deployment. To address these challenges, we propose and validate a CL–based framework for domain-incremental KWS that is suitable for embedded devices. The framework employs a compact classifier with only 1,595 parameters and uses MFCC and Log-Mel spectrogram features (20×16 for 1-second audio) as a lightweight rehearsal buffer. We first established a KWS pipeline suitable for embedded devices as it meets the storage and compute requirements for real time operation. To tackle the challenge of dynamic or varying deployed environment, we then developed an algorithm that utilizes the periodic retraining of the classifier model in our KWS pipeline using the combination of rehearsal buffer and effective samples determined at runtime. During CL phase, we retrain the model and update the class prototypes, decision thresholds for effective sample determination and the quantized model. To enable computation of robust audio features and ensure that only reliable features are fed into the classifier model for classification, our proposed denoising algorithm includes two stages, one before feature extraction and one after feature extraction. The model performance on various noisy conditions validate robust operation as the worst performance being observed at -10dB for NFIELD environment and the best performance at 10 dB for TCAR environment.   

Currently, domain-incremental learning is focused on binary classification which satisfies only a subset of the edge usecases. Our next focus is on expanding the CL for domain-incremental learning for multi-class classification to handle datasets such as MNIST, CIFAR-10, KWS-35 and CIFAR-100.  

\balance{}

\end{document}